# Cylinder-Specific Model-Based Control of Combustion Phasing for Multiple-Cylinder Diesel Engines Operating with High Dilution and Boost Levels

Wenbo Sui, Carrie M. Hall, and Gina Kapadia

Illinois Institute of Technology, Department of Mechanical, Materials, and Aerospace Engineering

## Abstract:

Accurate control of combustion phasing is indispensable for diesel engines due to the strong impact of combustion timing on efficiency. In this work, a non-linear combustion phasing model is developed and integrated with a cylinder-specific model of intake gas. The combustion phasing model uses a knock integral model, a burn duration model and a Wiebe function to predict CA50 (the crank angle at which 50% of the mass of fuel has burned). Meanwhile, the intake gas property model predicts the EGR fraction and the in-cylinder pressure and temperature at intake valve closing (IVC) for different cylinders. As such, cylinder-to-cylinder variation of the pressure and temperature at intake valves closing is also considered in this model. This combined model is simplified for controller design and validated. Based on these models, two combustion phasing control strategies are explored. The first is an adaptive controller that is designed for closed-loop control and the second is a feedforward model-based control strategy for open-loop control. These two control approaches were tested in simulations for all six cylinders and the results demonstrate that the CA50 can reach steady state conditions within 10 cycles. In addition, the steady state errors are less than ±0.1 crank angle degree (CAD) with the adaptive control approach, and less than ±1.3 CAD with feedforward model-based control. The impact of errors on the control algorithms is also discussed in the paper.

## Introduction

Concern for fuel consumption and pollutant output from the transportation sector has driven a recent surge in research on advanced internal combustion engines. On medium and heavy-duty ground vehicles, diesel engines are the primary power generation method and a number of technologies have been implemented on these engines in recent years in order to improve their efficiency and cleanliness. These technologies include turbochargers that are used to compress fresh air and enable higher power production from a smaller engine as well as exhaust gas recirculation (EGR) in which a portion of the exhaust flow is mixed with the fresh air to reduce nitrogen oxide emissions. Figure 1 shows a modern diesel engine with a turbocharger and exhaust gas recirculation.

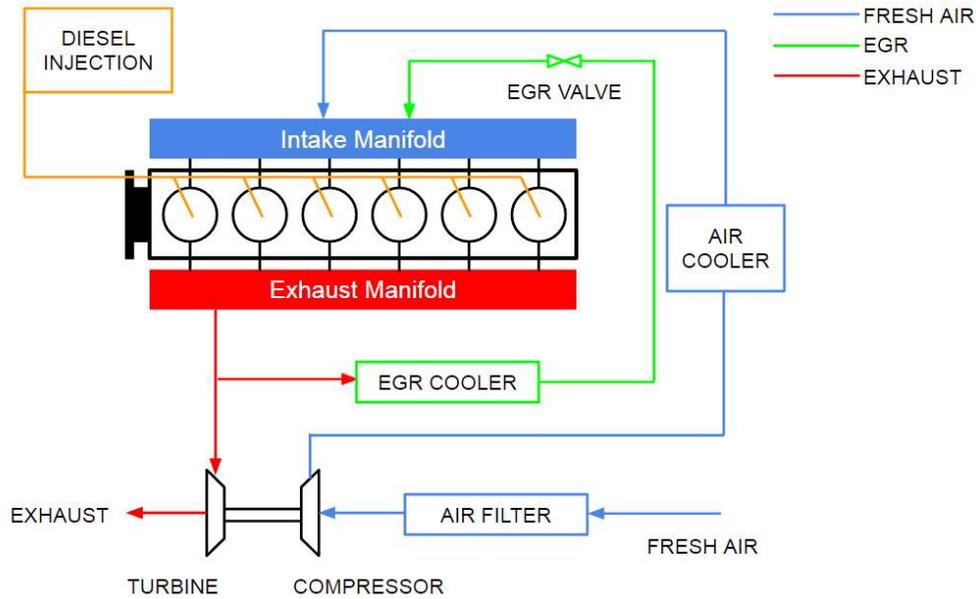

Figure 1.  Schematic of diesel engine system

While these technologies can improve the fuel efficiency and reduce the emissions of modern diesel engines, they also create a more complex system that requires more sophisticated control methodologies. Rule-based methodologies and look-up tables [1, 2] are now being replaced with model-based approaches and closed loop control strategies that rely on additional sensors. Model-based approaches have been leveraged to control fresh air flow, EGR flow, and combustion phasing. In this work, a model-based strategy is used for combustion control on a diesel engine that is operating with high boost levels and high dilution rates. While operating with higher boost and dilution can enable more efficient and clean operation, control of combustion phasing can be more challenging in such conditions since increase cylinder-to-cylinder variations can occur [23]. Furthermore, maintaining an optimal combustion phasing during engine transients is critical for ensuring that high efficiencies are maintained [3-6]. Combustion timing estimation strategies that use additional measurements of torque have also been proposed [7, 8], but are not commonplace since torque sensors are not standard on today's engines. In contrast, this paper studies a model-based method for controlling combustion phasing that is achievable with the sensors available on production engines.

Combustion has been modeled using a variety of methods. Arrhenius models have often been used to predict the start of combustion (SOC) [9-12]. While these models can be accurate, they often rely on variables such as in-cylinder oxygen fraction or fuel concentrations, which are not measured on stock engines. Other models have been proposed that use CFD models to tune their parameters properly [13-15]. Again, such models are able to predict the SOC, but may not be as useful due to the need for a detailed CFD model.

Another common way to capture the SOC is by using a knock integral model or KIM. While the KIM was originally used to predict knock on spark-ignited engines [16], it has also been leveraged more recently to predict SOC [17-19] as well as CO and NO emissions [18-19] on diesel engines. KIMs have also be applied to more advanced combustion strategies including homogenous charge compression ignition (HCCI) engines [20-21].

The authors have previously combined a KIM with a Wiebe function to predict CA50 for diesel engines and shown that this method can predict CA50 within ±0.5 CAD [22]. While this strategy worked, it assumed that the EGR fraction and pressure and temperature at intake valve closing (IVC) were known. However, such measurements are rarely made on production engines and in fact cylinder-to-cylinder variations on engines using high dilution and boost can be significant [23]. Therefore, the intake gas properties may need to be estimated for real applications on production engines. In [24] and [25], Al-Durra et al. and Stockar et al. have employed a sliding mode observer and a model-order reduction method to predict the in-cylinder pressure. Although these methods can predict the pressure at IVC accurately, they disregard heat transfer and are more complex which makes these methods more challenging for use in CA50 controller design. Chen et al. devised an extended Kalman filter-based method to estimate the in-cylinder temperature at IVC [26]. This method can estimate the temperature at IVC very accurately, but again the application of this method on production engines is challenging due to its complexity. A semi-empirical model was used to predict the pressure and temperature at IVC for HCCI engines [20]. This model's predictions were based on the engine speed, fuel equivalence ratio, EGR fraction, residual gas fraction and temperature as well as the intake manifold pressure and temperature. It also used an iterative method to estimate the residual gas temperature based on a non-linear model of that variable [20]. In contrast, this work examines a similar but simpler model of the cylinder-specific intake gas properties and integrates this with the authors' previously developed combustion model.

With an accurate model, a variety of model-based control methods can be pursued for combustion phasing control. Model-based control of CA50 (the crank angle when 50% of the fuel mass has burned) was studied in [17], but the controller took over 20 cycles to settle to steady state. Feedback control of CA50 and indicated mean effective pressure (IMEP) has also been examined in [27] and [28], but required in-cylinder pressure sensors. Similarly control of CA50 using feedback from crankshaft torque measurements [29] and control of the location of peak premixed combustion (LPPC) from an ion current signal [30] have also been shown to perform well. However, such sensors are not standard on production engines so the application of these methods may be limited and feedforward control techniques may be more useful.

Adaptive feedforward control using neural networks has been used to control the start of combustion and been shown to settle to steady state much faster that methods that used traditional PI feedback control [31]. Other feedforward strategies have shown promising results, but required iterative procedures such as that in [32]. Similarly, CA50 was controlled on an HCCI engine using feedforward control with an online adaptive estimator in [33].

While various studies have examined different models of the intake gas properties or combustion phasing and a variety of control methods have been introduced, this study integrates simple models with two control approaches should be appropriate for high efficiency, production diesel engines. Combustion phasing will be captured by the CA50 because combustion efficiency is tied to an optimal combustion phasing [6, 34]. While some prior studies have examined control of SOC or LPPC, these are not as strongly tied to fuel efficiency. This study combines an EGR fraction model and semi-empirical IVC pressure and temperature models with a modified KIM to predict SOC, Wiebe function to calculate CA50 based on the predicted SOC and a separate burn duration model. The integrated model is non-linear and as such is simplified for control purposes. It is calibrated with more than 200 simulations and validated against a detailed simulation model. This study considers a single injection only; multi-pulse injections or rate shaping were outside the scope of this study.

The main contributions of this study are 1) the creation of an integrated model of intake gas properties and combustion phasing to provide cylinder-specific CA50 predictions, 2) the calibration and validation of this model in conditions that are of interest for high efficiency diesel engines, namely, high EGR rates and high boost pressures, 3) simplification of these models for control applications, 4) consideration of the performance of this model in two different control methodologies, both open and closed loop. The controllers evaluated in this work are an adaptive feedback controller that leverages the non-linear model and a measurement of CA50. While some production diesel engines have this sort of feedback, most do not. To accommodate this, an open-loop, feedforward model-based method is also studied.

This paper first discusses the models of the intake gas properties and combustion phasing that are leveraged in this work as well as the calibration and validation of the models. Then, it focuses on two different control strategies, one open loop and one closed loop. The results of these control approaches are discussed and the impact of sensor errors on the controllers' performances are also considered.

# Modeling Diesel Engine Combustion Phasing

A control-oriented model of combustion phasing was created and it will be leveraged in the next section for control. In particular, this model predicts the CA50. Fig. 2 shows a block diagram of the CA50 prediction model. This model uses an EGR fraction that is calculated based on the oxygen fractions at the intake manifold and exhaust manifold. Meanwhile, the pressure and temperature at IVC are derived by semi-empirical models in the "Intake Gas Properties Model". SOC and BD are predicted by a modified KIM and burn duration model, respectively. Then, CA50 is calculated by a Wiebe function. The details of each model are discussed in this section.

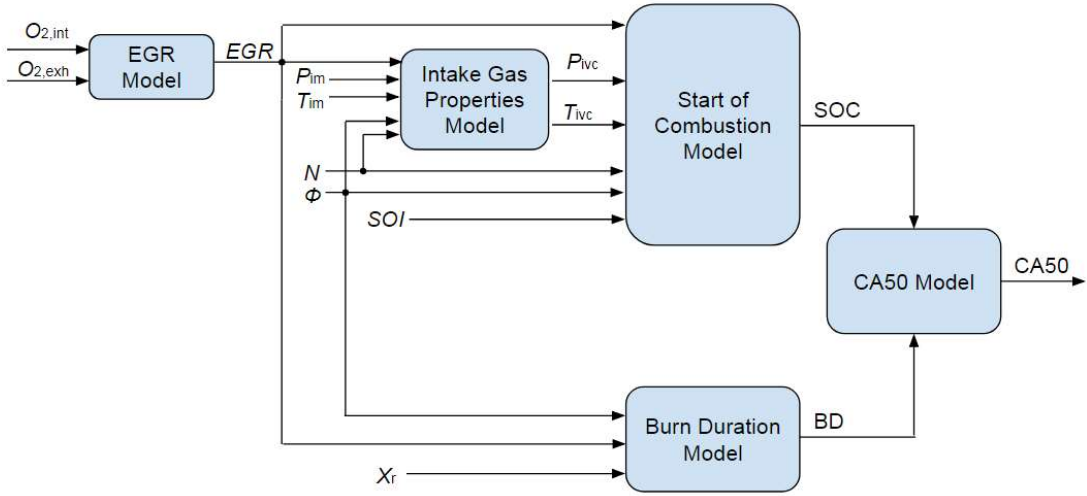

Figure2. Block diagram of CA50 prediction model

# EGR Calculation

EGR fraction is often predicted using the measured oxygen fraction in the intake and exhaust manifolds [20, 35]. As such, the EGR fraction is estimated by

$$EGR = \frac{x_{O2,amb} - x_{O2,int}}{x_{O2,amb} - x_{O2,exh}} \quad (1)$$

in this work. While oxygen fraction measurements are leveraged in this work, EGR can also be predicted using other estimations techniques including those described in [36] and [37].

# Intake Gas Properties Models

Using the estimated EGR fraction, additional intake gas properties can be modeled. These properties are critical to accurately predicting the combustion phasing particularly on engines running with high dilution levels and boost. Here a semi-empirical model is used to capture the expected pressure and temperature at IVC for each cylinder. Cylinder-to-cylinder variations can be more significant for highly boosted and dilute operation so capturing cylinder specific properties may be essential.

Semi-empirical models have been used for similar applications in the past. For example, the model of temperature at IVC given in Eqn. (2) was developed by Shahbakhti [20] and applied to a RCCI engine in [35].

$$T_{IVC} = X_r T_r \left(\frac{c_{v,r}}{c_{v,t}}\right) + (1 - X_r)(c_1 T_{im}^2 + c_2 T_{im} + c_3)\frac{\phi^{c_4} N^{c_5}}{(1 + EGR)^{c_6}} \left(\frac{c_{v,ch}}{c_{v,t}}\right) \quad (2)$$

In this expression, $T_{IVC}$ is the in-cylinder temperature at IVC, $X_r$ is the in-cylinder residual gas fraction, $T_r$ is the temperature of this residual gas and $T_{im}$ is the intake manifold temperature. The terms $c_{v,r}$, $c_{v,t}$, and $c_{v,ch}$ are the specific heat at constant volume of the in-cylinder residual gas, the total in-cylinder gas, and the charge at the intake manifold, respectively. $\phi$ is the equivalence ratio of the fuel; $N$ denotes the engine speed; $EGR$ represents EGR fraction; and $c_1, c_2, c_3, c_4, c_5$, and $c_6$ are all constant coefficients.

While this model is effective, the residual gas temperature, $T_r$, is challenging to predict. The work in [20] and [35] requires several complex non-linear iterations to accurately predict this variable. As such, this model is more difficult to use in real-time control efforts. In this work, this model is simplified to make it more amenable to control. In the conditions studied in this work, $X_r$ is around 0.03 to 0.09. Therefore, the second term on the right-hand side in Eqn. (2) is likely to dominate and the first term can be dropped. While this will cause some loss of accuracy, it will make the model more useful for control efforts. The resulting IVC temperature model is

$$T_{IVC} = (c_1 T_{im}^2 + c_2 T_{im} + c_3) \frac{\phi^{c_4} N^{c_5} P_{im}^{c_7}}{(1 + EGR)^{c_6}} \tag{3}$$

where $P_{im}$ is the intake manifold pressure and $c_7$ is another constant coefficient.

The combustion phasing is highly dependent on the temperature as well as the pressure at IVC. Here, another semi-empirical model is used to describe $P_{IVC}$ as

$$P_{IVC} = T_{im}^{c_8} N^{c_9} P_{im} \tag{4}$$

This model is the same as that used in [20] and [35]. Eqs. (3) and (4) capture the critical intake gas properties that will be leveraged in the combustion predictions. CA50 is predicted using a SOC and burn duration model. These models are discussed in more detail in [22] but are discussed briefly here for completion.

## SOC Model

The start of combustion or SOC is predicted using a knock integral model (KIM) [16-20]. KIMs typically are of the form,

$$\int_{SOI}^{SOC} \frac{\tau}{N} d\theta = 1 \tag{5}$$

in which $SOI$ is the crank angle at the start of injection, $SOC$ is the start of combustion, $N$ is the engine speed, and $\tau$ is an Arrhenius function. In the model used here, the Arrhenius function used is

$$\tau = \frac{1}{c_{10} EGR + c_{11}} \phi^{c_{12}} \exp\left(-\frac{c_{13} P^{c_{14}}}{T}\right) \tag{6}$$

where $EGR$ represents the EGR fraction, $\phi$ is the fuel equivalence ratio, $T$ is the temperature, $P$ is pressure, and $c_{10}$, $c_{11}$, $c_{12}$, $c_{13}$ and $c_{14}$ are constants.

## Burn Duration Model

The CA50 model also requires a prediction of the burn duration. In this work, burn duration is defined as the period between CA10 and CA90 and is captured by

$$BD = c_{15}(1 + X_d)^{c_{16}} \phi^{c_{17}} \qquad (7)$$

where $X_d$ is the dilution fraction, $\phi$ is the equivalence ratio, and $c_{15}$, $c_{16}$ and $c_{17}$ are constant parameters. The dilution fraction includes both the EGR fraction and residual fraction, $X_r$, which is defined as

$$X_r = \frac{m_r}{m_{air} + m_{fuel} + m_{egr}}. \qquad (8)$$

In Eq. (8), $m_r$ is the mass of residual gas, $m_{air}$ is the mass of fresh air, $m_{fuel}$ is mass of the injected fuel and $m_{egr}$ is the mass of EGR gas entering the cylinder.

## CA50 Prediction Model

Once the intake gas properties, SOC, and BD are found, CA50 can be estimated using a Wiebe function [3, 4], which has the form

$$x_b(\theta) = 1 - \exp\left(-a\left[\frac{\theta - SOC}{BD}\right]^b\right) \qquad (9)$$

where $x_b$ is the mass fraction of burned fuel, and $a$ and $b$ are constant coefficients. Using Eqn. (9), CA50 can be determined by finding the crank angle at which $x_b$ is equal to 0.5. This leads to the following expression for CA50

$$CA50 = SOC + \left[\frac{\ln 2}{a}\right]^{1/b} BD \qquad (10)$$

which is derived in more detail in [22].

## CA50 Model Simplification

The CA50 prediction was further simplified in order to eliminate the integral term in Eqn. (5). Since the equivalence ratio, engine speed and EGR fraction are constant for a cycle, they can be removed from the integral yielding

$$\frac{\phi^{c_{12}}}{(c_{10} EGR + c_{11})N} \int_{SOI}^{SOC} \exp\left[-\frac{c_{13} P^{c_{14}}}{T}\right] d\theta = 1. \qquad (11)$$

However, the change in volume from SOI to SOC is also relatively small and as such, the temperature and pressure do not change dramatically from SOI to SOC. Eqn. (11) can then be further simplified by using the pressure and temperature at SOI which can be known a priori in place of the actual dynamic pressure and temperature. This additional simplification allows Eqn. (11) to be reduced to the expression

$$\frac{\phi^{c_{12}}}{(c_{10}EGR + c_{11})N} \exp\left(-\frac{c_{13}P_{SOI}^{c_{14}}}{T_{SOI}}\right)(SOC - SOI) = 1. \tag{12}$$

Eqn. (12) can be rearranged and combined with Eqn. (10) to give the following expression for CA50:

$$CA50 = SOI + (c_{10}EGR + c_{11})N\phi^{-c_{12}} \exp\left(\frac{c_{13}P_{SOI}^{c_{14}}}{T_{SOI}}\right) \\ + c_{18}(1 + X_d)^{c_{16}}\phi^{c_{17}}. \tag{13}$$

Thus, the intake gas properties can be captured by Eqns. (1), (3), and (4) and CA50 can be estimated by Eqn. (13) using the EGR fraction, engine speed, equivalence ratio, pressure and temperature at SOI, and dilution fraction.

# CA50 Model Validation

## Diesel Engine Simulation Setup

The CA50 prediction model developed in the prior section must be calibrated and validated before using it for control efforts. A simulation model based on a 2010 Navistar Maxxforce 13 heavy-duty engine was used in the calibration and validation effort. The engine is a modern 6 cylinder 12.4L diesel engine that has a variable geometry turbocharger (VGT) and cooled EGR. The specifications of this engine are listed in Table 1.

Table 1. Engine Specifications

| | |
|---|---|
| Displacement Volume | 12.4L |
| Number of Cylinders | 6 |
| Compression Ratio | 17:1 |
| Valves per cylinder | 4 |
| Bore | 126mm |
| Stroke | 166mm |
| Connecting Rod length | 251mm |
| Diesel Fuel System | 2200 bar common rail |
| Air System | 2-stage turbocharger |

Experiments on this testbed were conducted by Kassa and are discussed in more detail in [23]. A simulation model of the air path system of this engine was also developed in [23].

Based on the experimental data, a combustion simulation model in GT-ISE (Gamma Technology Integrated Simulation Environment) was calibrated and validated by Hulbert based on three pressure analysis (TPA) method [38]. This TPA model can predict CA50 in the simulation with a ±0.84 CAD uncertainty as compared to experimental results [38]. Additional details regarding the agreement between the GT-ISE simulation model and experimental results can be found in [22] and [38]. Prior work has shown that the cylinder-to-cylinder variations observed on this engine and are mainly caused by uneven mixing between the EGR, the quantity of fresh air, and any port-injected fuel. These variations were captured in the GT-ISE model by using a detailed gas exchange model [23]. Because the fuels used and the cylinder geometry are the same for all cylinders, the combustion model in each cylinder is the same in the GT-ISE model.

In this work, 288 simulations were tested to achieve enough data for calibration and validation. The range of the parameters is shown in Table 2. While this range of data does not cover the whole engine operating range, it does cover a range in which high dilution rates and boost can be leveraged to achieve high efficiencies. As such, this is a region of interest.

Table 2. Range of Parameters in Simulation Data

| Quantity | Minimum Value | Maximum Value |
| --- | --- | --- |
| Engine Speed (RPM) | 1200 | 1500 |
| Average Intake Manifold Temperature (K) | 302.52 | 333.29 |
| Average Intake Manifold Pressure (bar) | 1.43 | 2.97 |
| Diesel Equivalence Ratio (-) | 0.5 | 0.9 |
| EGR (%) | 0 | 50 |
| SOI (° aTDC) | -10 | 0 |
| IVO (° aTDC) | -363.5 | -363.5 |
| IVC (° aTDC) | -148.5 | -148.5 |
| EVO (° aTDC) | 137 | 137 |
| EVC (° aTDC) | 389 | 389 |

EGR: exhaust gas recycling; SOI: fuel start of injection; aTDC: after top dead center; IVO: intake valve open; IVC: intake valve close; EVO: exhaust valve open; EVC: exhaust valve close.

# Calibration and Validation of CA50 Prediction Model

In order to calibrate the models discussed in the previous section, the models of the intake gas properties in Eqns. (3) and (4) were calibrated as well as the CA50 model in Eqn. (13). Because each cylinder may have different intake gas properties, the coefficients in Eqns. (3) and (4) need to be calibrated separately. Unlike the intake gas properties models, the CA50 prediction model leveraged the same parameters for different cylinders.

The calibration process is shown in Fig. 3. The root mean squared error (RMSE) of both sides in Eqns. (3), (4), and (13) is set as the reference in the calibration process. The RMSE is minimized based on a batch gradient descend algorithm. Initial conditions were set at the beginning of the calibration process. Afterwards, the initial value of the RMSE gradient is set to zero. The errors are computed for each simulation and then the error gradient is calculated. These gradients are added to the gradient of RMSEs. After the calculation of all 288 simulations, the coefficients are updated based on the gradient of RMSEs. The calibration iterations will stop when the RMSE cannot decrease.

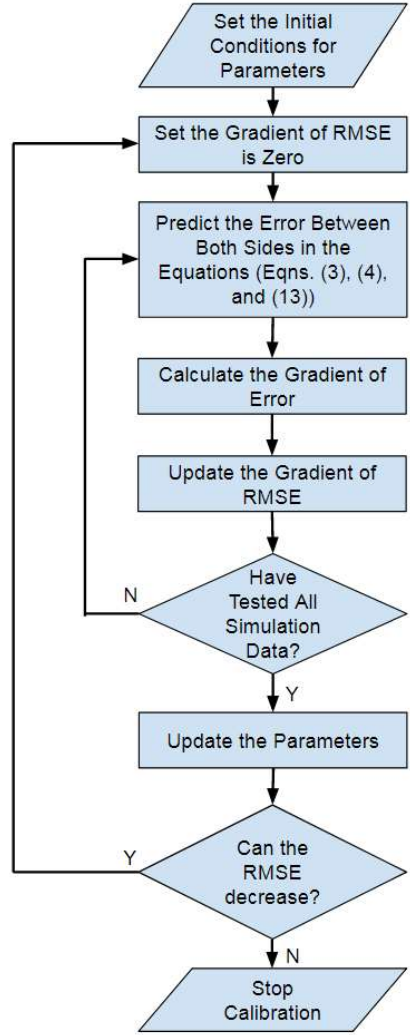

Figure 3.     Model calibration procedure

The calibrated coefficients are listed in Tables 3-5. Table 3 and Table 4 show the parameters for Eqns. (3) and (4) for all six cylinders and the optimized parameters for the CA50 model in Eqn. (13) are given in Table 5.

Table 3.     Parameters of the $T_{IVC}$ Prediction Model

| Cylinder Number | 1 | 2 | 3 | 4 | 5 | 6 |
|---|---|---|---|---|---|---|
| $c_1$ | $-7.35 \times 10^{-4}$ | $-7.68 \times 10^{-4}$ | $-8.00 \times 10^{-4}$ | $-8.24 \times 10^{-4}$ | $-8.53 \times 10^{-4}$ | $-8.73 \times 10^{-4}$ |
| $c_2$ | 0.842 | 0.855 | 0.850 | 0.852 | 0.844 | 0.839 |
| $c_3$ | -12.1 | -10.2 | -11.2 | -5.21 | -4.29 | 4.69 |
| $c_4$ | 0.111 | 0.109 | 0.114 | 0.112 | 0.116 | 0.112 |
| $c_5$ | -0.167 | -0.165 | -0.161 | -0.168 | -0.165 | -0.166 |
| $c_6$ | 0.0204 | 0.0195 | 0.0177 | 0.0171 | 0.0152 | 0.0136 |
| $c_7$ | 0.0600 | 0.0602 | 0.0585 | 0.0594 | 0.0598 | 0.0594 |

Table 4.     Parameters of the $P_{IVC}$ Prediction Model

| Cylinder Number | 1 | 2 | 3 | 4 | 5 | 6 |
|---|---|---|---|---|---|---|
| $c_8$ | -0.0580 | -0.0504 | -0.0462 | -0.0488 | -0.0386 | -0.0388 |
| $c_9$ | 0.0810 | 0.0713 | 0.0651 | 0.0710 | 0.0569 | 0.0582 |

Table 5.     Parameters of CA50 Prediction Model

| | |
|---|---|
| $c_{10}$ | $1.11 \times 10^{-5}$ |
| $c_{11}$ | $8.03 \times 10^{-4}$ |
| $c_{12}$ | $7.56 \times 10^{-2}$ |
| $c_{13}$ | $8.22 \times 10^{4}$ |
| $c_{14}$ | -1.15 |
| $c_{16}$ | 4.59 |
| $c_{17}$ | 0.628 |
| $c_{18}$ | 0.0251 |
| $k_c$ | 1.25 |

Substituting the parameters in Tables 3, 4, and 5 into Eqns. (3), (4) and (13), CA50 can be predicted. These predicted CA50s are compared with the more detailed simulation CA50s form the GT-ISE model, which were used to calibrate the model. The comparisons between the model predicted SOC and CA50 and that from GT-ISE are shown in Fig. 4 and Fig. 5, respectively. Statistics on the prediction accuracy are given in Table 6. In these figures, the x axis is the value from the GT-ISE simulation, and the y axis is the predicted value from the calibrated control-oriented model. Perfect agreement corresponds to the solid blue line, while the dashed blue lines represent the conditions with ±1 CAD errors.

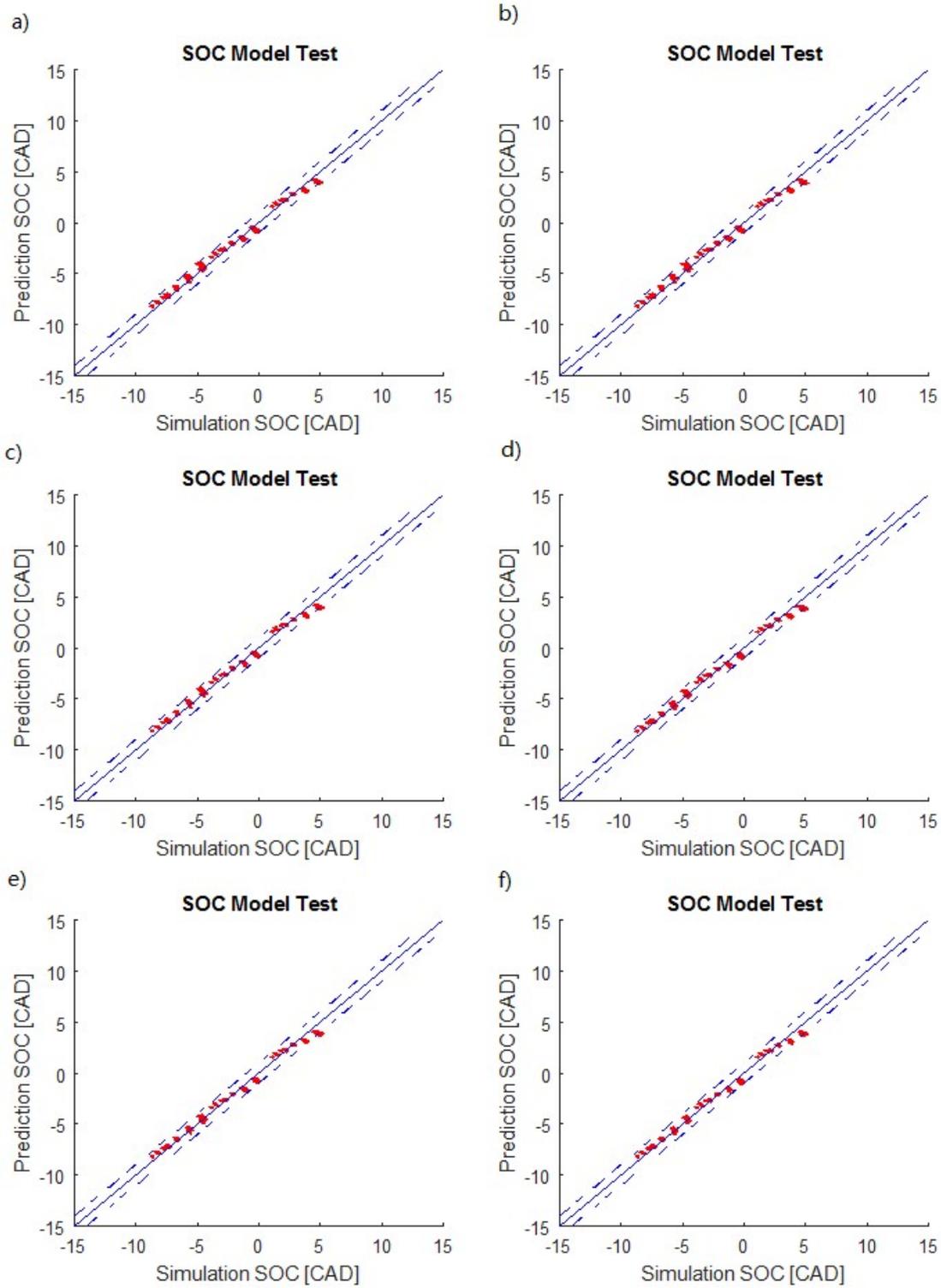

Figure 4. Comparison of SOC prediction and simulation data: a) cylinder 1, b) cylinder 2, c) cylinder 3, d) cylinder 4, e) cylinder 5, and f) cylinder 6

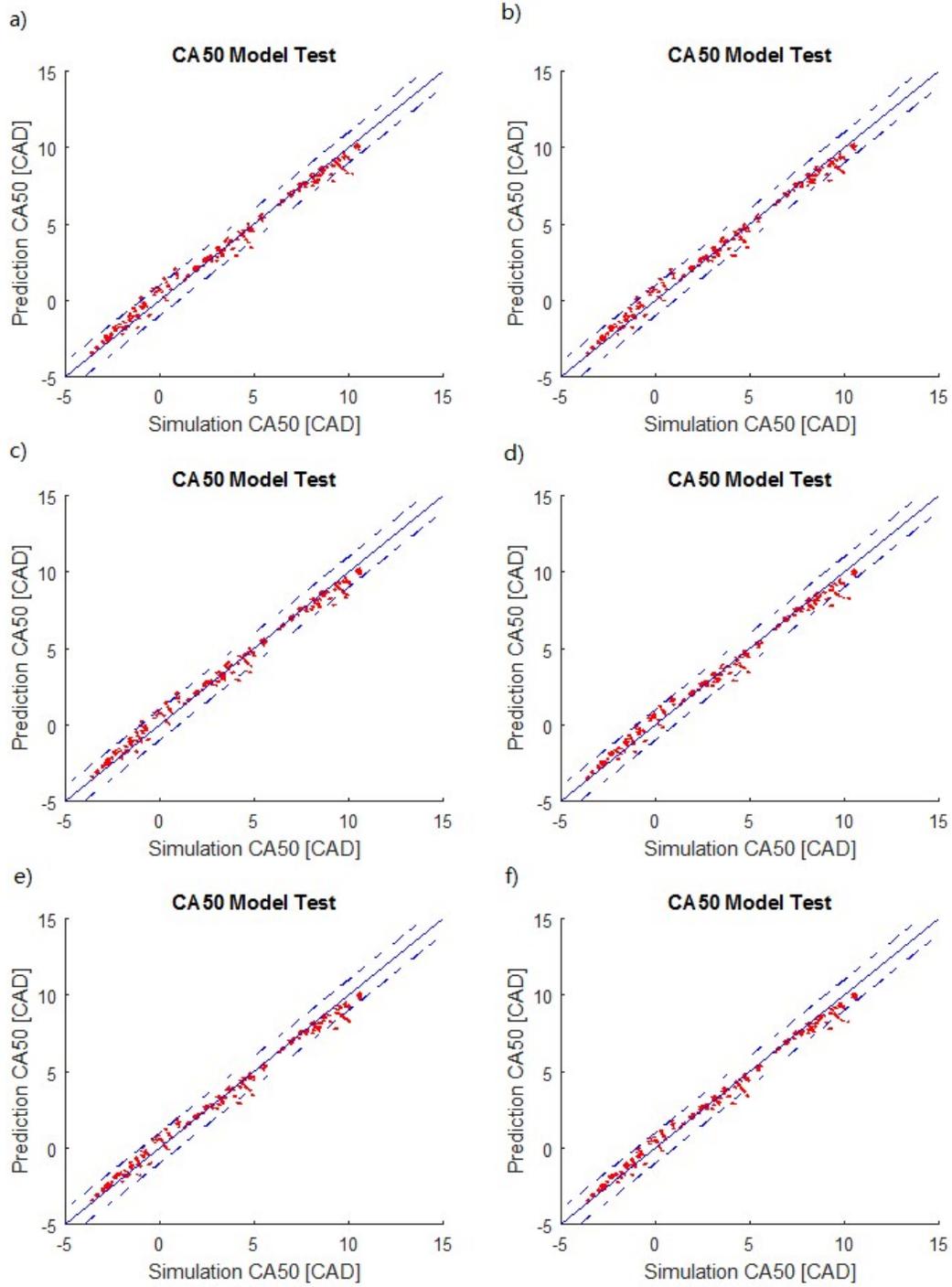

Figure 5. Comparison of CA50 prediction and simulation data: a) cylinder 1, b) cylinder 2, c) cylinder 3, d) cylinder 4, e) cylinder 5, and f) cylinder 6

Table 6. SOC and CA50 Prediction Model Validation Result

| Cylinder Number | 1 | 2 | 3 | 4 | 5 | 6 |
|---|---|---|---|---|---|---|
| Standard Deviation of SOC Prediction Error (CAD) | 0.43 | 0.43 | 0.44 | 0.44 | 0.44 | 0.44 |
| Maximum SOC Prediction Error (CAD) | 1.28 | 1.31 | 1.34 | 1.37 | 1.39 | 1.39 |
| Standard Deviation of CA50 Prediction Error (CAD) | 0.48 | 0.48 | 0.47 | 0.47 | 0.46 | 0.45 |
| Maximum CA50 Prediction Error (CAD) | 1.93 | 1.69 | 1.59 | 2.00 | 1.75 | 1.96 |

Comparing the prediction error of SOC and CA50, most of the CA50 prediction error is from the errors in the prediction of SOC. However, the maximum prediction error of CA50 is higher than that of SOC. CA50 prediction is much worse at high EGR fractions and a more accurate prediction model of the burn duration may need to be explored in future. Based on the comparison in Fig. 4, Fig. 5, and Table 6, it can be seen that the CA50 prediction model can calculate the CA50 with high accuracy in all six cylinders. The standard deviations of the prediction errors are less than 0.5 CAD, and the maximum values of the errors are no more than 2.0 CAD. Therefore, this CA50 model should be able to be leveraged for model-based control of combustion phasing in multiple-cylinder diesel engines. In the next section, two model-based control strategies are developed. The first is an adaptive controller that is suitable for closed loop systems and the second is an open loop control method.

## Adaptive Control Strategy Design

Although the CA50 prediction model is expressed in Eqn. (13), it is necessary to transform into state-space form for the adaptive controller design. In this system, CA50 is set as the output of system $y$, and SOI is the input of system $u$, which is used to track the reference CA50 values. The exponential term and the dilution term are employed as the states in the state-space model. Thus, the state-space equations of the CA50 prediction model can be captured by following equations:

$$y = u + \alpha x_1 + \beta x_2 \qquad (14)$$

$$x_1 = (c_{10}EGR + c_{11}) \exp\left(\frac{c_{13}P_{SOI}{}^{c_{14}}}{T_{SOI}}\right) \tag{15}$$

$$x_2 = c_{18}(1 + X_d)^{c_{16}} \tag{16}$$

$$\alpha = N\phi^{-c_{12}} \tag{17}$$

$$\beta = \phi^{c_{17}} \tag{18}$$

where $EGR$ represents EGR fraction, $P_{SOI}$ and $T_{SOI}$ are the temperature and pressure at SOI, $X_d$ denotes the dilution fraction, $N$ indicates the engine speed, $\phi$ is the equivalence ratio, and $c_{10}$ - $c_{18}$ are the parameters shown in Table 5. In the state-space version of the CA50 dynamic model, the states $x_1$ and $x_2$ vary between different cycles and cylinders, and the coefficients $\alpha$ and $\beta$ are derived with the measurements of engine speed $N$ and equivalence ratio $\phi$ as in Eqns. (17) and (18).

## Adaptive Control System Structure

Using the state space model, a model-based adaptive controller can be developed. Figure 6 shows a block diagram of the CA50 adaptive feedback control system. In this control strategy, the reference CA50 is set as the optimal CA50. This reference signal along with the measurements of engine speed and the equivalence ratio of fuel are sent to the adaptive controller. Besides these signals, feedback of the actual CA50 measurement also serves as an input signal to the controller. With these input signals, a suitable SOI is calculated, and this information is then sent to the diesel engine system.

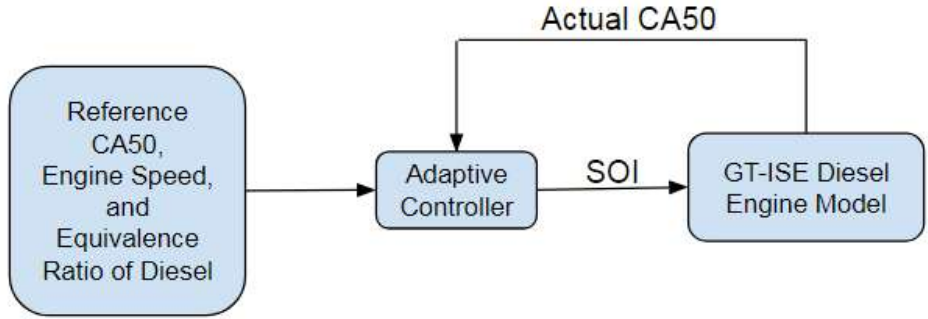

Figure 6. Structure of CA50 adaptive feedback control system

## Adaptive Controller Design

With the dynamic model in Eqn. (14), the input of the control system $u$ can be derived as

$$u = y_d - \alpha x_1 - \beta x_2 \tag{19}$$

in which $y_d$ represents the desired output of the control system, $\alpha$ and $\beta$ are the coefficients given by Eqns. (17) and (18). Unlike these coefficients, the states $x_1$ and $x_2$ vary from cycle to cycle, and are not typically measured by sensors. Therefore, an observer is used to update the values of these states. Replacing the states by their estimated values, Eqn. (19) can be written as

$$u = y_d - \alpha \bar{x}_1 - \beta \bar{x}_2 \tag{20}$$

where $\bar{x}_1$ and $\bar{x}_2$ are the states estimations.

To update the values of the states, a gradient descent algorithm is utilized. In this work, the error function is defined as the squared error or

$$E = \frac{1}{2}(y - \bar{y})^2 \tag{21}$$

where $E$ indicates the RMSE between the actual CA50 and the observed CA50, $y$ expresses the actual CA50 and $\bar{y}$ is the observed CA50. The observed CA50 $\bar{y}$ is given by:

$$\bar{y} = u + \alpha \bar{x}_1 + \beta \bar{x}_2. \tag{22}$$

Rewriting Eqn. (20), the desired output can be written as:

$$y_d = u + \alpha \bar{x}_1 + \beta \bar{x}_2. \tag{23}$$

Because the parameters and the states are invariant when the system achieves steady state, it can be found that the observed output $\bar{y}$ equals the desired output $y_d$ from Eqns. (22) and (23). Substituting $\bar{y}$ by $y_d$ in Eqn. (21), the error function is

$$E = \frac{1}{2}(y - y_d)^2. \tag{24}$$

Based on the error function expressed in Eqn. (24), the partial derivatives of the RMSE can be calculated by:

$$\frac{\partial E}{\partial \bar{x}_1} = -\alpha(y - y_d) \tag{25}$$

$$\frac{\partial E}{\partial \bar{x}_2} = -\beta(y - y_d) \tag{26}$$

With these partial derivatives, the observed states can be updated from cycle to cycle in a gradient descent algorithm by

$$\bar{x}_1(k+1) = \bar{x}_1(k) - \eta \frac{\partial E}{\partial \bar{x}_1} \tag{27}$$

$$\bar{x}_2(k+1) = \bar{x}_2(k) - \eta \frac{\partial E}{\partial \bar{x}_2} \tag{28}$$

in which $\bar{x}_1(k)$, $\bar{x}_2(k)$, $\bar{x}_1(k+1)$, and $\bar{x}_2(k+1)$ denote the observed states at $k$ cycle and $k+1$ cycle, respectively; $\eta$ indicates the learning rate of the gradient descent algorithm; and $\frac{\partial E}{\partial \bar{x}_1}$ and $\frac{\partial E}{\partial \bar{x}_2}$ are the partial derivatives expressed in Eqns. (25) and (26).

Substituting the partial derivatives in Eqns. (25) and (26) into Eqns. (27) and (28), the observed states are then given by Eqns. (29) and (30) below.

$$\bar{x}_1(k+1) = \bar{x}_1(k) + \eta \alpha(y - y_d) \tag{29}$$

$$\bar{x}_2(k+1) = \bar{x}_2(k) + \eta \beta(y - y_d) \tag{30}$$

In order to keep the control system stable and achieve a short settling time and small overshoot, the learning rate $\eta$ can be calculated by

$$\eta = \frac{0.3}{\alpha^2 + \beta^2}. \tag{31}$$

Substituting Eqn. (31) into Eqns. (29) and (30), the update equations of the observer are:

$$\bar{x}_1(k+1) = \bar{x}_1(k) + \frac{0.3\alpha}{\alpha^2 + \beta^2}(y - y_d) \tag{32}$$

$$\bar{x}_2(k+1) = \bar{x}_2(k) + \frac{0.3\beta}{\alpha^2 + \beta^2}(y - y_d) \tag{33}$$

With Eqns. (20), (32), and (33), the control system can track CA50 with its reference value. A proof of the stability of this adaptive control system stability is given in the Appendix.

# Feedforward Model-Based Controller Design

Because a CA50 signal may always be reliable during transient conditions or not even available on production diesel engines, a feedforward model-based controller without this feedback measurement may be needed. To achieve this goal, a controller based on a feedforward control strategy is designed, and its block diagram is shown in Fig. 7.

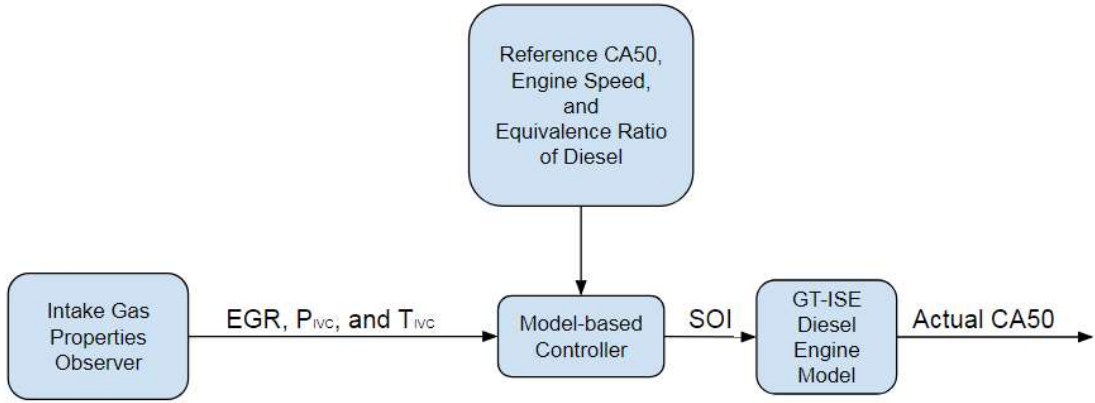

Figure 7.  Structure of CA50 feedforward model-based control system

As in the control system shown in Fig. 7, the intake gas properties including EGR fraction, $P_{IVC}$ and $T_{IVC}$, are predicted by Eqns. (1), (3), and (4). The reference CA50 is given by an optimal CA50 and the model-based controller is able to calculate the appropriate SOI based on the intake gas properties, engine speed, and the equivalence ratio.

Changing the form of Eqn. (13), the SOI can be computed by:

$$SOI = CA50_{\text{ref}} - (c_{10}EGR + c_{11})N\phi^{-c_{12}} \exp\left(\frac{c_{13}P_{SOI}{}^{c_{14}}}{T_{SOI}}\right) \\ - c_{18}(1 + X_d)^{c_{16}}\phi^{c_{17}} \tag{34}$$

where $CA50_{\text{ref}}$ is the reference CA50.

The dilution fraction, the pressure and temperature at SOI in Eqn. (34) are not available from measurements directly. The $P_{SOI}$ and $T_{SOI}$ can be captured by a polytropic correlation. Therefore, the start of injection can be given by:

$$SOI = CA50_{\text{ref}} - (c_{10}EGR + c_{11})N\phi^{-c_{12}} \exp\left[\frac{c_{13}\left[P_{\text{IVC}}\left(\frac{V_{\text{IVC}}}{V_{\text{SOI}}}\right)^{k_c}\right]^{c_{14}}}{T_{\text{IVC}}\left(\frac{V_{\text{IVC}}}{V_{\text{SOI}}}\right)^{k_c-1}}\right] \quad (35)$$

$$-c_{18}(1 + EGR + X_r)^{c_{16}}\phi^{c_{17}}$$

in which $V_{\text{SOI}}$ denotes the volume of cylinder at SOI.

To calculate SOI, the parameters in Eqn. (35) are required. The EGR fraction is calculated using the oxygen fraction in the intake and exhaust manifold as in Eqn. (1). The engine control unit (ECU) can provide the engine speed $N$ directly and the equivalence ratio of diesel can be calculated based on the commanded air and fuel flows. $P_{\text{IVC}}$ and $T_{\text{IVC}}$ are predicted by Eqns. (3) and (4). The volume of cylinder at IVC $V_{\text{IVC}}$ is determined by the geometry of the cylinder. True $V_{\text{SOI}}$ prediction would require several iterations of Eqn. (35), which is too complicated for real applications. Because the SOI does not change dramatically, the SOI from the last cycle is used to calculate the volume of cylinder at SOI in this study. The residual fraction $X_r$ only ranges from 0.0344 to 0.0909 in all simulations and it is much smaller than the EGR fraction. Therefore, the mean value of the residual fraction $\overline{X_r}$ ($\overline{X_r} = 0.0642$) is employed to predict SOI in feedforward control strategy.

## Simulation and Analysis

In previous section, both an adaptive controller and feedforward controller are introduced. To evaluate the performance of these control strategies, both controllers are tested in four different cases. These cases include:
1. A change in boost pressure
2. A change in reference CA50
3. A change in engine speed and equivalence ratio
4. A change in engine speed, equivalence ratio and EGR fraction

In these cases, the estimates of the engine temperature and pressure begin at ambient values. The engine stays at one operating condition for 10 seconds and as such, the estimates stabilize over the first engine cycles. Then the operating condition changes to a second set of conditions from 10 to 20 seconds. The resulting CA50 are plotted for these cases and are the predicted CA50s from the GT-ISE model. The details of the settings along with the simulation results and analysis are shown in each case. In Case 1, the simulation results of adaptive controller and feedforward controller are compared with the performance of PID controller. The SOI in cylinder 1 is also discussed to demonstrate how SOIs in two control systems are influenced by the change of boost pressure.

### Performance During Boost Pressure Change

First, the controllers are tested during a change in the target boost pressure. The settings for this case are listed in Table 7.

Table 7. Settings of Case 1

| Quantity | First Operating Point | Second Operating Point |
|---|---|---|
| Engine Speed (RPM) | 1200 | 1200 |
| Average Charge Temperature (K) | 300 | 300 |
| Target Boost Pressure (bar) | 1.5 | 2.5 |
| Diesel Equivalence Ratio (-) | 0.7 | 0.7 |
| EGR fraction (%) | 25 | 25 |
| Reference CA50 (CAD) | 8 | 8 |

The performance of the adaptive controller is shown in Fig. 8, while the feedforward controller's performance is in Fig. 9. The actual CA50s in all cylinders along with the reference CA50 are labeled in the figures. In the first cycle, fuel is not injected into the cylinders, and no combustion occurs. The fuel is injected into the cylinders beginning at the second cycle, and the controllers begin to work in the third cycle. Because there is no active CA50 control in the second cycle, the CA50s have a large overshoot in both control systems. Due to the firing order, the CA50s in cylinder 2 and cylinder 4 are delayed. As in Fig. 8, the actual CA50s in the adaptive control system have large oscillations until they reach steady state, but the steady state error is only -0.04 to 0.06 CAD after 5 seconds. This oscillation is highly related to the significant changes in the average intake manifold pressure as shown in Fig. 11. The average intake manifold pressure jumps from its initial condition 1 bar to 2.2 bar in only 1.3 second. Then it drops down and oscillates until the 6 second mark. The fast changes in intake manifold pressure lead to the rapid changes in the states of the adaptive controller. Because the adaptive controller tracks the states from the last cycle, the observation of the states has a large error when they are changing rapidly. Therefore, the actual CA50s have oscillations until the intake manifold pressure reaches its steady state. Although this degrades the performance of the adaptive controller and may be more excessive than would actually occur experimentally, the adaptive control system has CA50s errors less than ±1 CAD after 10 cycles.

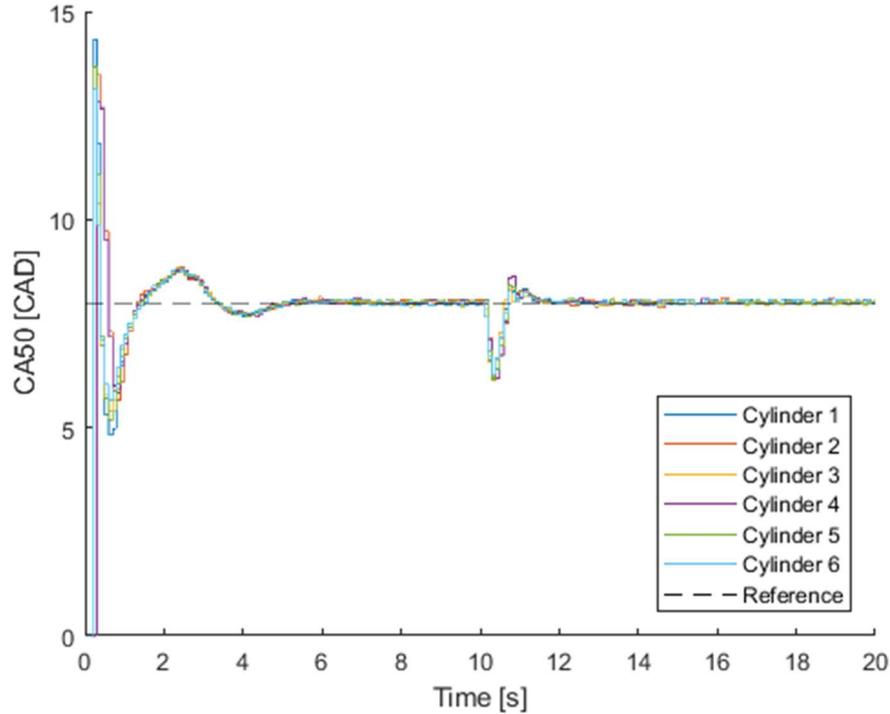

Figure 8.    Adaptive Control Simulation Result for Case 1

Since the feedforward strategy uses a different method to calculate SOI, the CA50s in the feedforward control system do not have such a phenomenon. The CA50s in the feedforward control system have errors of less than 0.6 CAD before the change in operating condition. After 10 seconds, the operating condition changes, and the target boost pressure rises from 1.5 bar to 2.5 bar. The feedforward controller has steady state errors from 0.07 to 0.26 CAD in the first operating condition, and has steady state errors from 0.08 to 0.27 CAD in the second operating condition.

A PID controller was also tested in Case 1 and the simulation result for a PID control system is plotted in Fig. 10. The CA50s in this control system reach their steady state at 7.6 seconds in the first operating condition with the steady state errors less than ±0.1 CAD. Because the boost pressure oscillates significantly before steady state, the PID controller cannot track the desired CA50s during this oscillation. The same situation also happens after the change of boost pressure. The CA50s track the desired reference 15 cycles later than the boost pressure reaches its steady state. In short, the PID can achieve its steady state with steady state error less than ±0.1 CAD. However, the performance of the PID controller during transients is poor.

In contrast, both the adaptive and feedforward controllers achieve steady state in 10 cycles. Overall, the adaptive controller has steady state errors less than ±0.1 CAD, which are lower than the feedforward controller as expected. According to the simulation results shown in Figs. 8 and 9, both the adaptive controller and feedforward controller perform well for all cylinders, and they can achieve their steady states in 10 cycles with ±1 CAD errors. Compared with the adaptive control strategy, the feedforward control has better performance when the boost pressure has rapid changes, but it has more steady state error.

Compared with PID controller, these two controllers have a much better performance during transients, and the adaptive controller has similar performance with the PID controller at the steady state, which is better than the feedforward controller.

While both of the more advanced controllers perform well, they do differ. The average intake manifold pressure is plotted in Fig. 11, and the SOIs that are commanded by the adaptive controller and feedforward controller are given in Fig. 12. Compared with the trend of intake manifold pressure, it can be concluded that the SOIs delay when the pressure increases, and advance when the pressure decreases. The higher intake manifold pressure leads to higher pressure at SOI. Due to higher pressure at SOI, the second term in the right-hand side of Eqn. (13) decreases because $c_{14}$ is negative. In order to achieve the desired CA50, the SOI must delay. Thus, the SOIs have a similar trend as the intake manifold pressure. It also can be found in Fig. 12 that the SOI for the feedforward control system changes earlier than the SOI from the adaptive control system. The feedforward controller provides SOIs based on the measurements in last cycle, while the adaptive controller calculates the SOIs based on those update of parameters, which need several cycles to track their actual value. Because of this, the adaptive controller changes SOIs later than the feedforward controller, and it has worse performance during the transient.

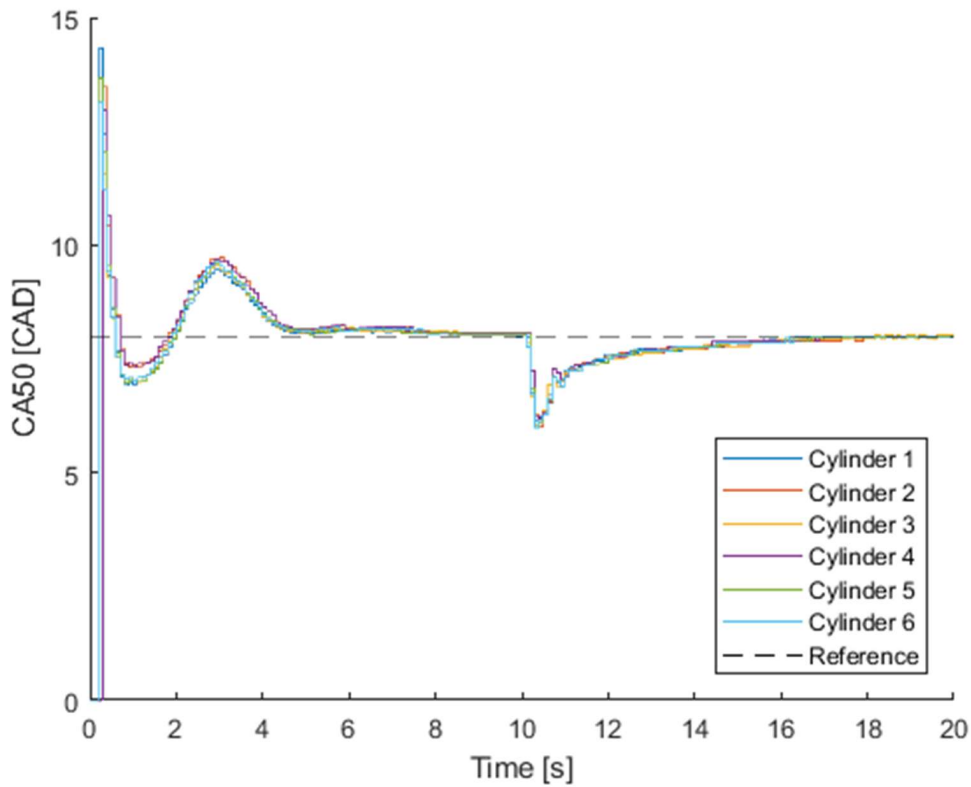

Figure 10. PID Control Simulation Result for Case 1

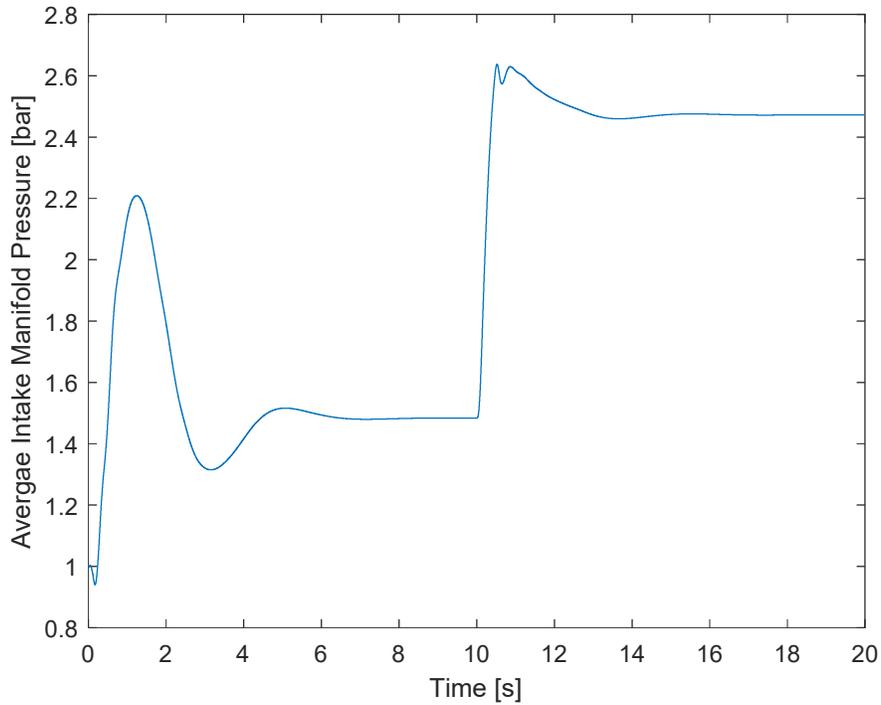

Figure 11.     Average Intake Manifold Pressure for Case 1

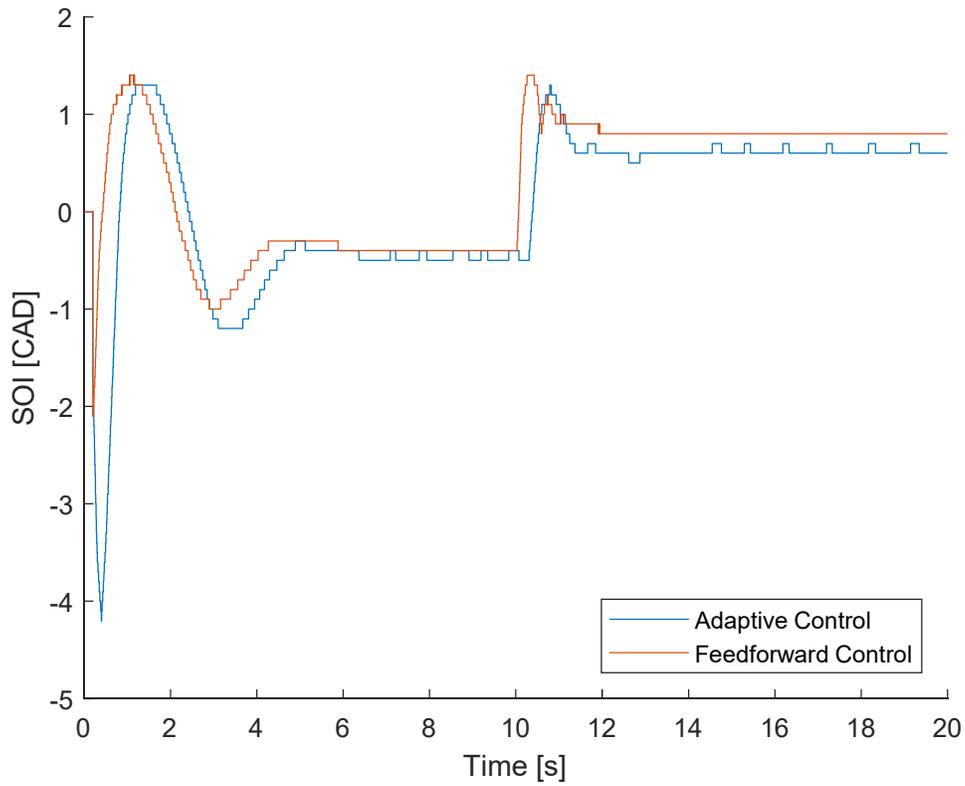

Figure 12.     SOI in Cylinder 1 for Case 1

## Performance During Reference CA50 Change

The controller performance during a change in the reference CA50 was evaluated in Case 2. The settings of this case are listed in Table 8, and the simulation results are plotted in Figs. 13 and 14.

Table 8. Settings of Case 2

| Quantity | First Operating Point | Second Operating Point |
| --- | --- | --- |
| Engine Speed (RPM) | 1200 | 1200 |
| Average Charge Temperature (K) | 300 | 300 |
| Target Boost Pressure (bar) | 2 | 2 |
| Diesel Equivalence Ratio (-) | 0.7 | 0.7 |
| EGR fraction (%) | 25 | 25 |
| Reference CA50 (CAD) | 8 | 10 |

As in Case 1, combustion does not occur in the first cycle, and the controllers are not active in the first 2 cycles. Also similar to Case 1, the actual CA50 with adaptive control has an oscillation, but it more quickly settles to steady state. The steady state errors in CA50 range from -0.04 to 0.06 CAD during the first operating condition. Compared with this adaptive control strategy, the system with feedforward controller again does not have such significant oscillations. The steady state error with feedforward control at the first operating condition ranges from -0.09 to 0.08 CAD. The reference CA50 jumps to 10 CAD at 10 seconds. After this change, the actual CA50 in both control systems increases as desired. Unlike the performance in feedforward control system, the actual CA50s in the adaptive control system have overshoots from 0.6 to 1.2 CAD. After the transient, the steady state errors are from -0.08 to 0.09 CAD with adaptive control and from -0.03 to 0.13 CAD with feedforward control.

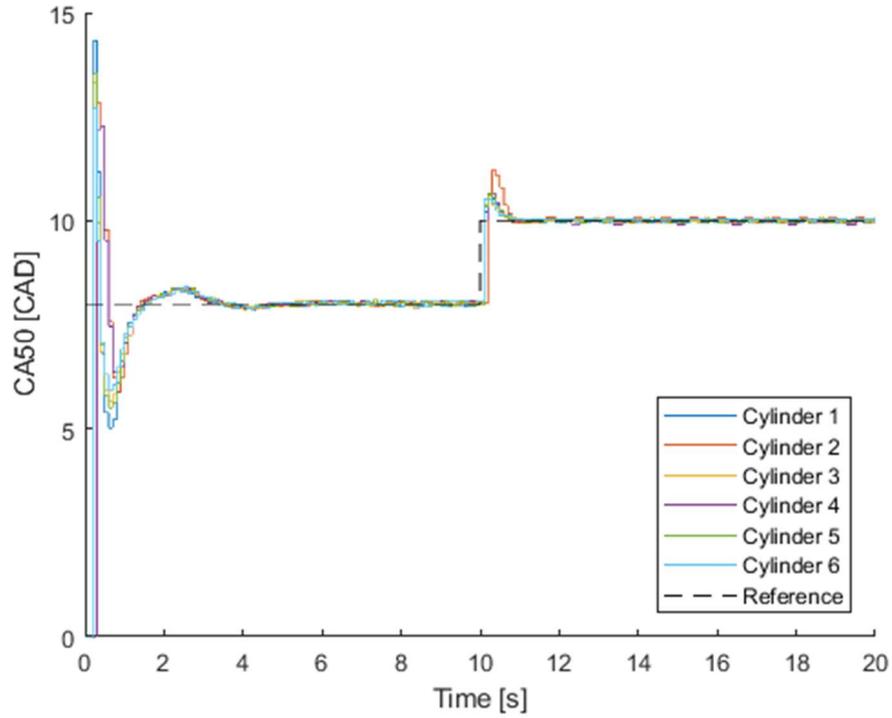

Figure 13.     Adaptive Control Simulation Result for Case 2

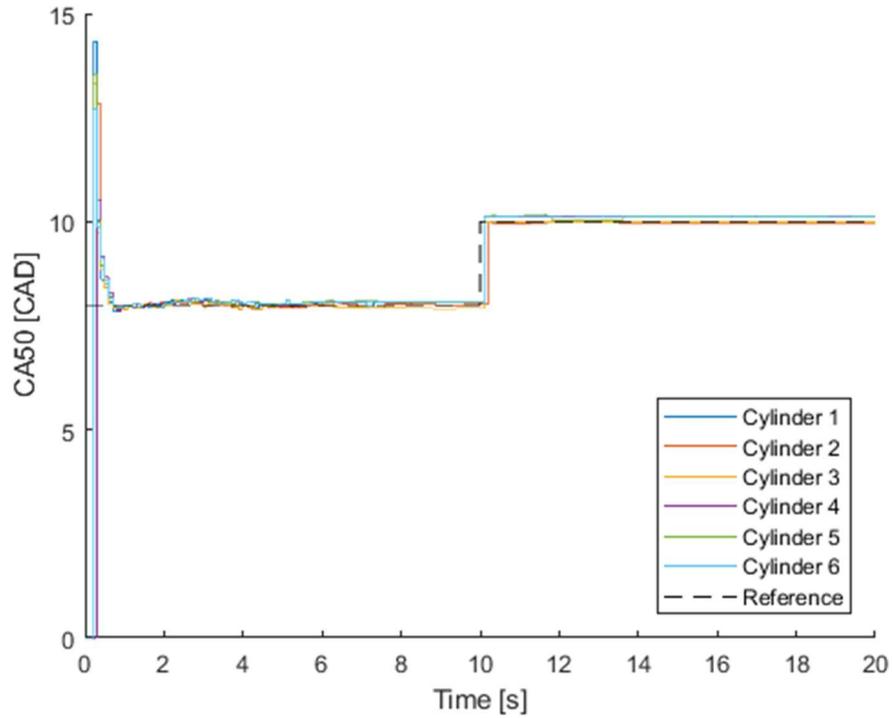

Figure 14.     Feedforward Control Simulation Result for Case 2

## Performance During a Combined Change in Equivalence Ratio and Engine Speed

The third case considers a transient in both the equivalence ratio and engine speed. In Table 9, the settings of this case are shown and the simulation results are also shown in Figs. 15 and 16. Again, the controllers start to work form the third cycle, and both controllers reach ±1 CAD error in 10 cycles.

Table 9. Settings of Case 3

| Quantity | First Operating Point | Second Operating Point |
|---|---|---|
| Engine Speed (RPM) | 1200 | 1500 |
| Average Charge Temperature (K) | 300 | 300 |
| Target Boost Pressure (bar) | 2 | 2 |
| Diesel Equivalence Ratio (-) | 0.5 | 0.9 |
| EGR fraction (%) | 25 | 25 |
| Reference CA50 (CAD) | 8 | 8 |

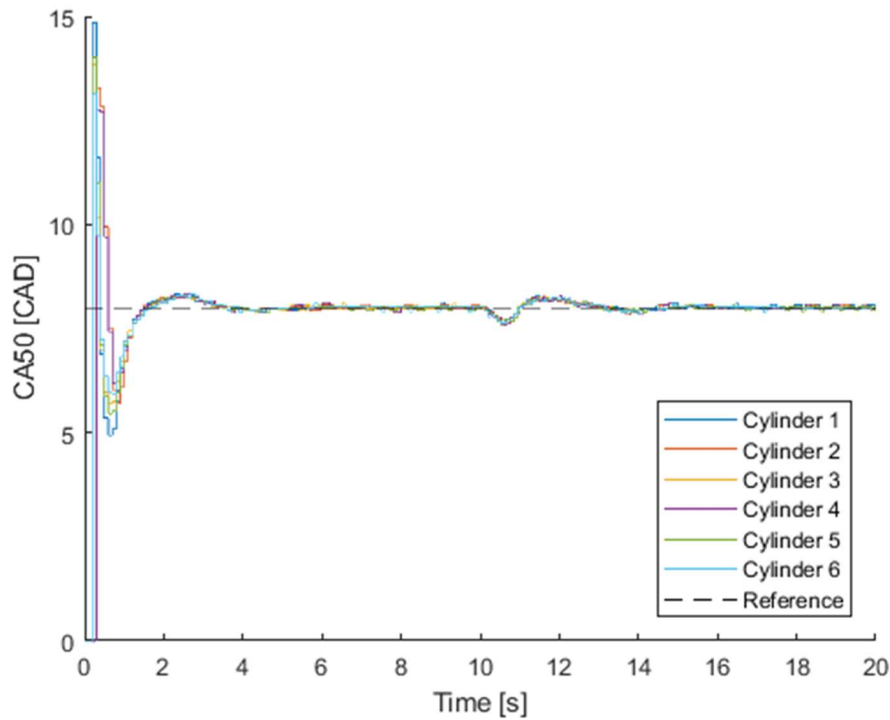

Figure 15. Adaptive Control Simulation Result for Case 3

Similar to Case 1 and Case 2, the steady state errors with adaptive control are from -0.08 to 0.03 CAD, and the errors with feedforward control are slightly higher and span from - 0.14 to 0.06 CAD. After 10 seconds, the equivalence ratio increases from 0.5 to 0.9 in 0.5 seconds, while the engine speed increases smoothly from 1200 RPM to 1500 RPM in 0.5 seconds. During the transient, the CA50 from the adaptive control has a slight oscillation with errors from -0.40 CAD to 0.26 CAD. This oscillation is due to large fluctuations in the boost pressure as demonstrated in Fig. 17. After the transient, the CA50s in both controllers come back to steady state. The steady state errors are from -0.04 to 0.08 CAD and from 0.21 to 0.30 CAD for the adaptive control system and the feedforward control system, respectively.

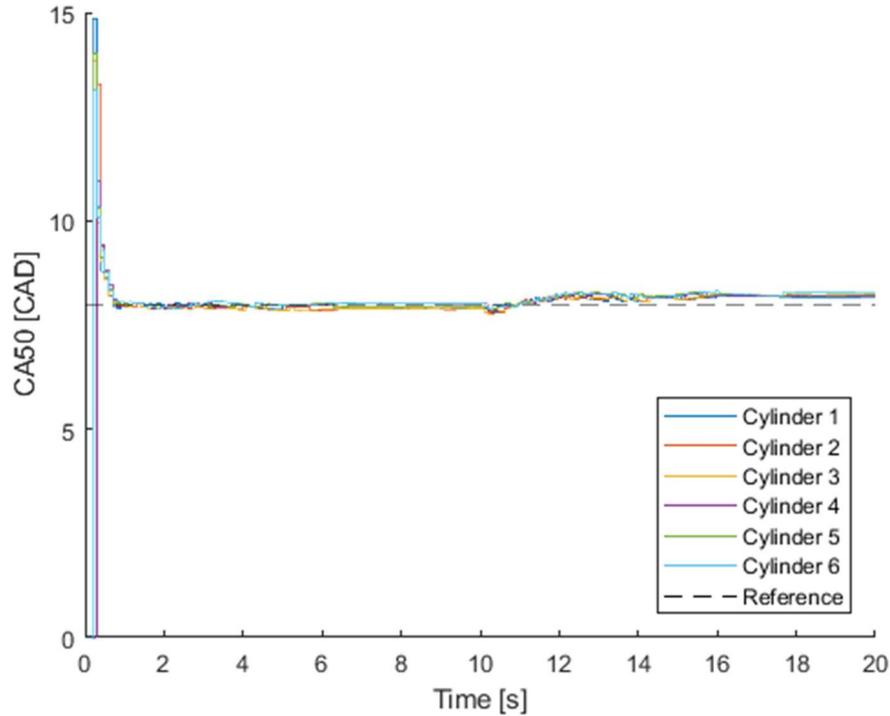

Figure 16.  Feedforward Control Simulation Result for Case 3

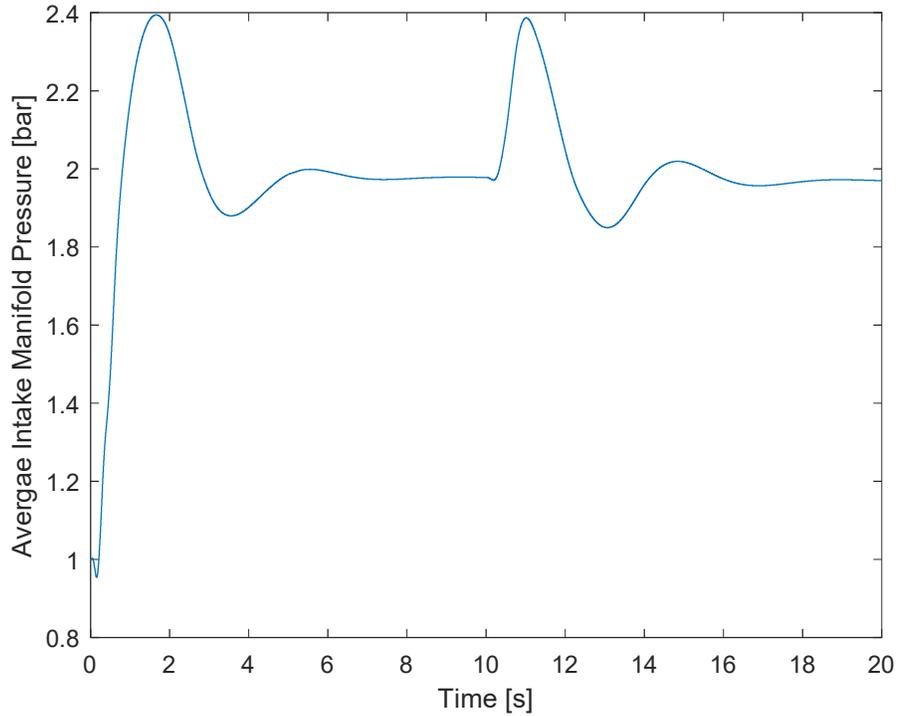

Figure 17.    Average Intake Manifold Pressure for Case 3

## Performance During a Combined Change in EGR Fraction, Equivalence Ratio, and Engine Speed

Lastly, a combined change in EGR fraction, equivalence ratio, and engine speed is tested using the settings given in Table 10.

Table 10.    Settings of Case 4

| Quantity | First Operating Point | Second Operating Point |
|---|---|---|
| Engine Speed (RPM) | 1200 | 1500 |
| Average Charge Temperature (K) | 300 | 300 |
| Target Boost Pressure (bar) | 2 | 2 |
| Diesel Equivalence Ratio (-) | 0.5 | 0.9 |
| EGR fraction (%) | 0 | 50 |
| Reference CA50 (CAD) | 8 | 8 |

The simulation results are shown in Figs. 18 and 19. As in the three cases above, the controllers do not work in the first two cycles. Similar to the other three cases, the CA50 oscillates until the 15th cycle due to the oscillation of the boost pressure. The steady state errors in the first operating condition are from -0.08 to 0.08 CAD with the adaptive controller and higher at 0.08 to 0.23 CAD with the feedforward controller. After 10 seconds, the equivalence ratio and engine speed smoothly increase in 0.5 second from the first operating condition to second operating condition, while the target of EGR fraction goes from 0% to 50%. Although the desired EGR fraction is changed in one step, the actual EGR fraction increases from 0% to 50% in 0.7 seconds. Because of the change in EGR fraction, equivalence ratio and the engine speed, the boost pressure also changes during the transient. This change of the boost pressure leads to an 0.70 to 0.75 CAD overshoot of CA50 in the adaptive control system. The steady state error in the adaptive control system is from -0.10 to 0.08 CAD. However, the steady state error in the feedforward control system is 0.94 to 1.21 CAD, which is much higher than the adaptive control system. This high steady state error is related to the error of the CA50 prediction model at the second operating condition. Compared with the simulation result in Case 3, higher steady state error is seen in the second operating condition in Case 4 due to the change in EGR fraction. When the EGR fraction is increased from 0 to 50%, the residual gas fraction changes from 0.0721 to 0.0415. However, in this open-loop control strategy, the residual fraction is kept constant at 0.0642. The difference between the dynamic residual gas fraction and the fixed residual gas fraction utilized in the controller leads to the error of the prediction of the third term in Eqn. (35). Dynamic residual gas fraction applied in the open-loop controller will be investigated in future work.

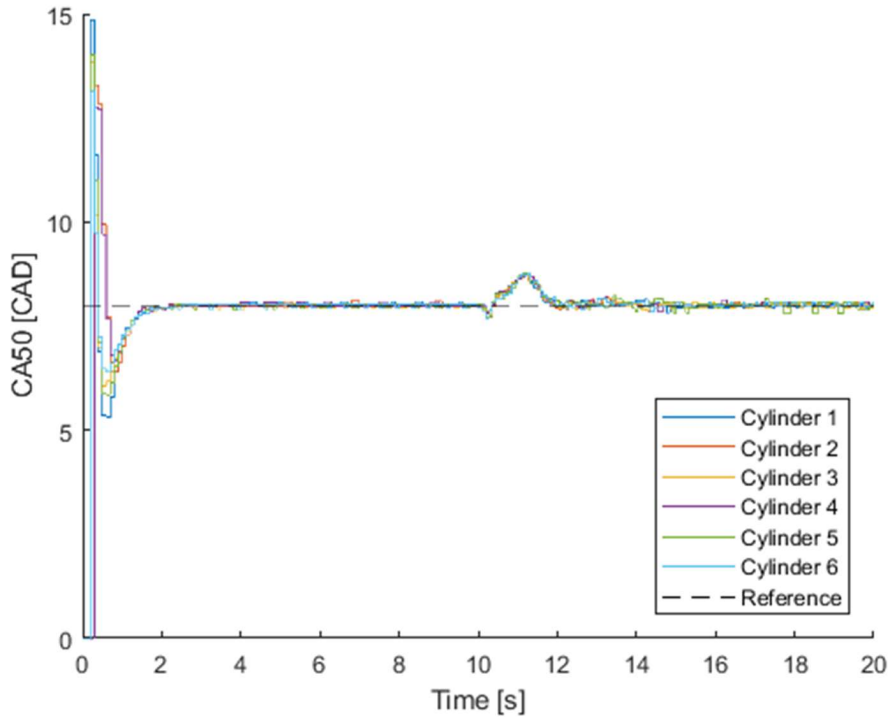

Figure 18.  Adaptive Control Simulation Result for Case 4

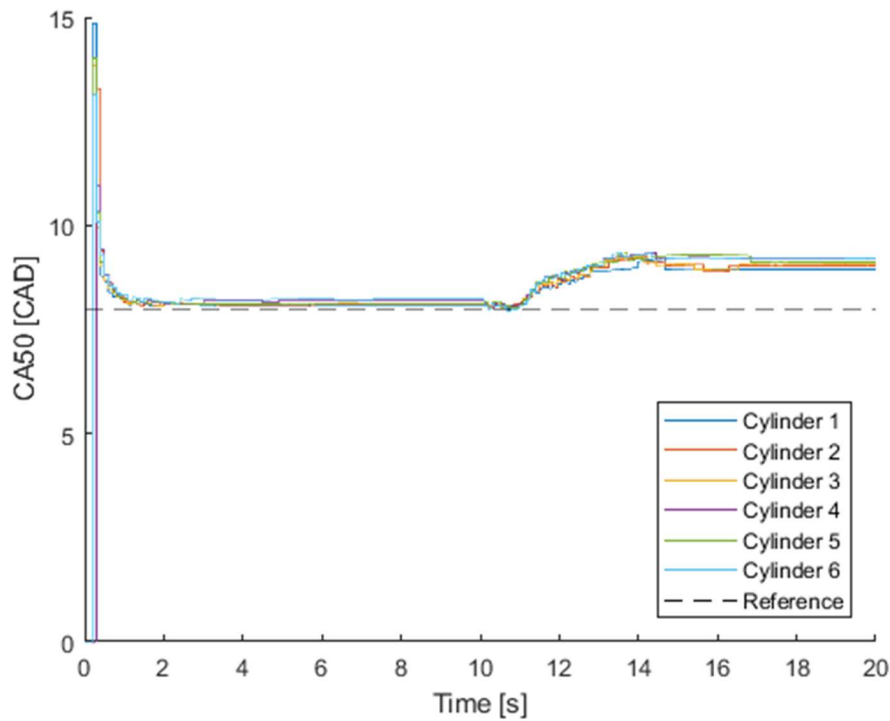

Figure 19.  Feedforward Control Simulation Result for Case 4

While the adaptive controller performs well, it should be noted that there are some errors that remain in the observed states, $x_1$ and $x_2$. A comparation between the observed states and the actual states has been plotted in Fig. 20 and Fig. 21. As seen in those figures, $x_1$ and $x_2$ can be observed with small steady state errors in the first operating condition. However, the observed $x_1$ and $x_2$ have much larger errors in the second operating condition. This is due to the fact that the observer leverages error in the CA50 prediction rather than error in the predicted state. As such, CA50 prediction is accurate, but the observed states can have errors that offset each other as seen in the second operating condition in Figures 20 and 21. Since the goal is accurate CA50 estimation, this is not a significant issue, but could be addressed in future work by examining different adaptation methods and learning rates.

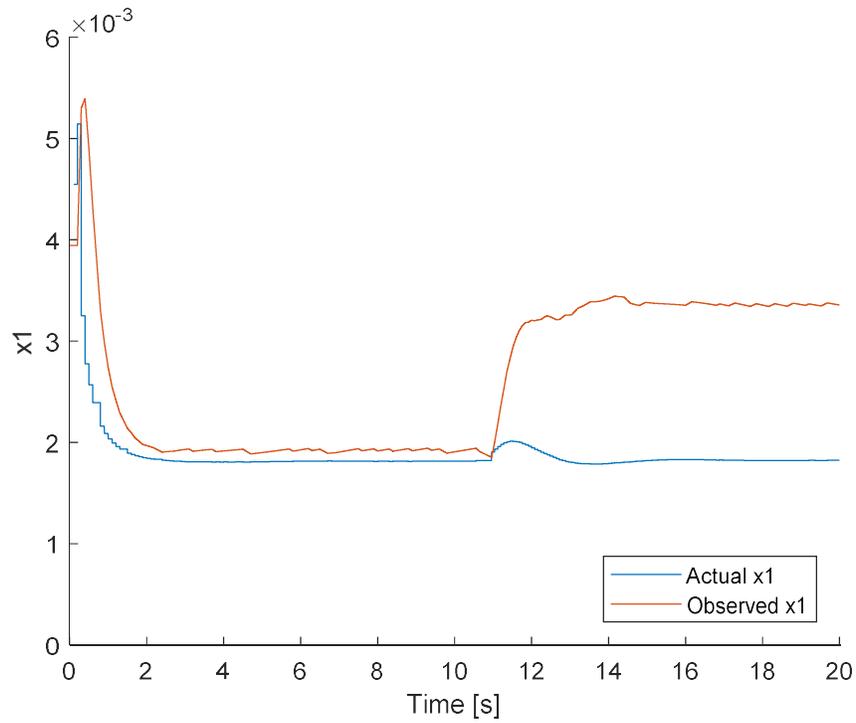

Figure 20. Observed vs. Actual $x_1$ in Cylinder 1 for Case 4

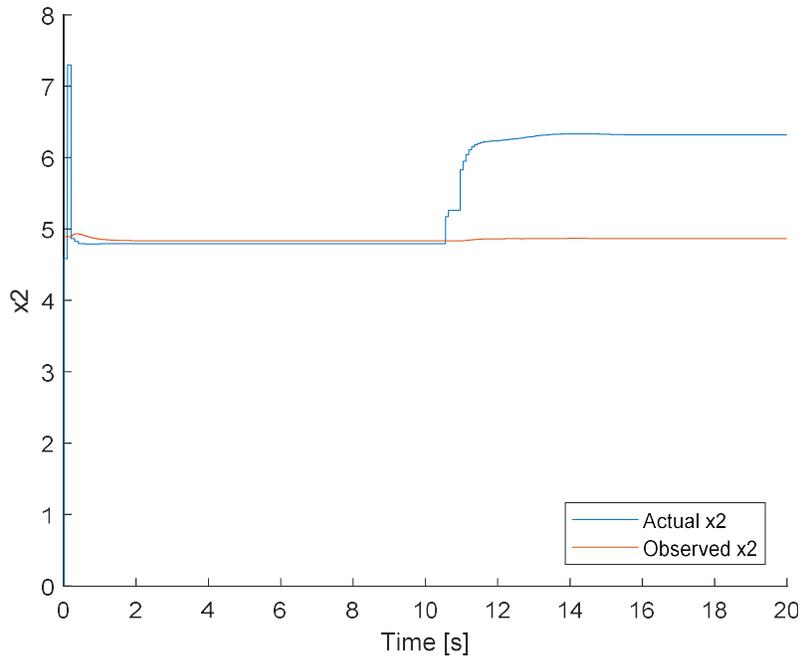

Figure 21. Observed vs. Actual $x_2$ in Cylinder 1 for Case 4

According to the simulation results, both control methods can track the reference CA50 quickly with a maximum steady state errors less than ±0.1 CAD for adaptive control, and

±1.3 CAD for feedforward control. Due to the delay in state updates with the adaptive controller, significant errors appear when the boost pressure changes quickly. Unlike the adaptive control strategy, the feedforward control system has better performance during the transient, but worse performance at the steady state. Because CA50 measurements or cylinder pressure feedback are not available for most production diesel engines today, the feedforward control strategy can be more widely applied in stock diesel engines.

## Discussion of Measurement Errors

While the performance of these controllers seems adequate, the impact of sensor measurement errors needs to be considered as well. First, the impact on the feedforward controller is discussed. For the feedforward controller investigated above, the control error is highly influenced by the prediction error of the model. Thus, the error response of the CA50 prediction model is tested. After that, the adaptive control system is also tested with measurement noise.

The error response of the CA50 prediction model could be influenced by several sources, such as the average temperature and pressure at the intake manifold, the EGR fraction, the equivalence ratio, and the residual gas fraction. The impact from errors in these variables were evaluated and the results are listed in Table 11. As shown in Table 11, the CA50 prediction model can calculate CA50 with a standard deviation less than 0.6 CAD, and a maximum error of less than 2.2 CAD. Therefore, the error from those measurements is still likely manageable for real-time applications.

Table 11. Error Response of CA50 Prediction

| Error Source | Error Value | Standard Deviation of CA50 Prediction Error | Maximum of CA50 Prediction Error |
|---|---|---|---|
| No Error | - | 0.47 | 2.00 |
| $T_{im}$ | +5 K | 0.46 | 2.02 |
| $T_{im}$ | -5 K | 0.48 | 1.99 |
| $P_{im}$ | +0.1 bar | 0.45 | 2.11 |
| $P_{im}$ | -0.1 bar | 0.58 | 2.17 |
| EGR | +5% | 0.48 | 1.99 |
| EGR | -5% | 0.47 | 2.02 |
| Equivalence Ratio | +0.05 | 0.47 | 2.00 |
| Equivalence Ratio | -0.05 | 0.48 | 2.01 |
| $X_r$ | +0.03 | 0.47 | 1.92 |
| $X_r$ | -0.03 | 0.47 | 2.08 |

In order to investigate the error response in the adaptive control system, the system was tested with CA50 measurement noise. The uncertainty of the CA50 measurements is set to 0.5 CAD, because the uncertainty of CA50 measurements from lab grade pressure sensors is 0.26 CAD. The simulation was run with the same settings as the first operating condition in Case 2. The simulation result in cylinder 1 is plotted in the Fig. 20, and the standard deviations of the errors during the simulation in all six cylinders are listed in Table 12. The standard deviation ranges from 1.29 to 1.61 CAD, because the transient in the first several cycles are included in the calculation. In fact, the steady state error is less than ±1 CAD. From Fig. 20 and Table 12, one can conclude that the adaptive controller can perform well with reasonable noise in the CA50 measurements.

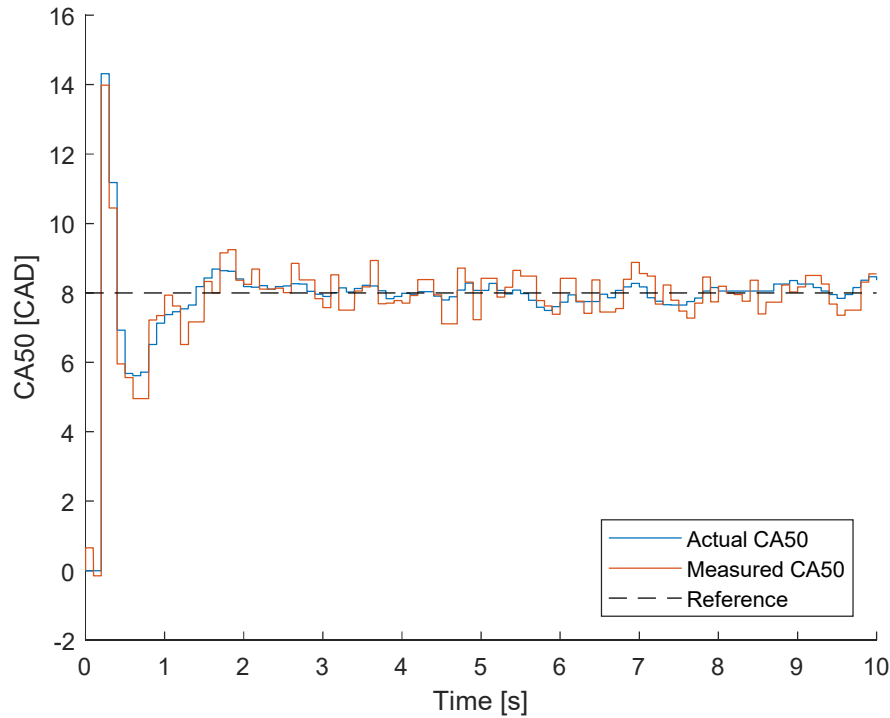

Figure 20. Simulation Result of Error Response of Adaptive Control System in Cyl. 1

Table 12.    Simulation Result of Error Response of Adaptive Control System

| Cylinder Number | Standard Deviation of Errors without CA50 Measurement Noise | Standard Deviation of Errors with CA50 Measurement Noise |
| --- | --- | --- |
| 1 | 1.4426 | 1.4218 |
| 2 | 1.5743 | 1.6151 |
| 3 | 1.3372 | 1.3720 |
| 4 | 1.5440 | 1.5815 |
| 5 | 1.3689 | 1.4094 |
| 6 | 1.2909 | 1.3248 |

# Conclusions

In this paper, a combustion phasing prediction model that includes prediction of the cylinder-specific intake gas properties is investigated for a diesel engine. These models are simplified, calibrated and validated for control design. In this study, the calibration and validation of the models is based on 288 simulations with a large range of boost pressures and EGR rates. A feedback control strategy and a feedforward control strategy are developed to track the optimal CA50. The adaptive controller is designed as a feedback controller that would use CA50 measurements (or cylinder pressure feedback) and has a steady state error below ±0.1 CAD at different operating conditions. Meanwhile, an open-loop controller is also discussed for the situations in which CA50 measurements are not available. For such a feedforward control system, the steady state error is less than ±1.3 CAD with the same operating conditions as used with the adaptive control system. Thus, both control systems can track the optimal combustion phasing with reasonable accuracy. In addition, the error introduced by sensor noise is also studied. According to the simulation results, the two control strategies still perform well when the sensor measurements have noise.

Because the methods studied in this paper are generalizable, the control strategies developed in this paper can be also leveraged for other compression ignition engines. In next steps, these control strategies will be evaluated experimentally and tested in drive cycles.

# Acknowledgement

This material is based upon work supported by the National Science Foundation under Grant No. 1553823.

# Nomenclature

| Coefficient | Definition |
|:---:|:---:|
| $BD$ | Crank angle during burn duration |
| $CA50$ | Crank angle at 50% of fuel mass burnt |
| $CA50_{ref}$ | Reference CA50 |
| $EGR$ | Exhaust Gas Recirculation Fraction |
| $k_c$ | Polytropic constant |
| $m_{air}$ | Air mass entering the cylinder |
| $m_{egr}$ | EGR mass entering the cylinder |
| $m_{fuel}$ | Mass of injected fuel |
| $m_{residual}$ | Mass of residual gas in the cylinder |
| $N$ | Engine Speed |
| $P$ | In-cylinder dynamic pressure |
| $P_{IVC}$ | Pressure at intake valve close |
| $P_{SOI}$ | Pressure at start of injection |
| $SOC$ | Crank angle at start of combustion |
| $SOI$ | Crank angle at start of fuel injection |
| $T$ | In-cylinder dynamic temperature |
| $T_{IVC}$ | Temperature at intake valve close |
| $T_{SOI}$ | Temperature at start of injection |
| $V$ | Dynamic volume of cylinder |
| $V_0$ | Cylinder volume at 0 crank angle degree |
| $V_{IVC}$ | Cylinder volume at intake valve close |
| $V_{SOI}$ | Cylinder volume at start of injection |
| $x_b$ | Mass fraction of burnt fuel |
| $X_r$ | Mass fraction of residual gas |
| $X_d$ | Mass fraction of dilution |
| $\tau$ | Arrhenius function |
| $\phi$ | Diesel equivalence ratio |
| $\theta$ | Crank angle |

# Acronyms

| | |
|---|---|
| AFR | Air fuel ratio |
| aTDC | After top dead center |
| BD | Burn duration |
| CA50 | Crank angle at 50% fuel mass burnt |
| CAD | Crank angle degree |
| CFD | Computational fluid dynamics |
| ECU | Engine control unit |
| EGR | Exhaust gas recirculation |
| EVC | Exhaust valve close |
| EVO | Exhaust valve open |
| HCCI | Homogenous charge compression ignition |
| IVC | Intake valve close |
| IVO | Intake valve open |
| KIM | Knock integral model |
| LPPC | Location of peak premixed combustion |
| MKIM | Modified knock integral model |
| PI | Proportional–integral |
| PID | Proportional–integral–derivative |
| RBFNN | Radial basis function neural network |
| RMSE | Root mean squared error |
| RPM | Revolutions per minute |
| SOC | Start of combustion |
| SOI | Start of injection |
| st | Stoichiometric |
| TPA | Three pressure analysis |
| VGT | Variable geometry turbocharger |
| VVT | Variable valve timing |

# Appendix: Proof of Adaptive Control System Stability

The stability of the adaptive control system is proved by the Lyapunov direct method. Since the purpose of the controller is to minimize the error between the desired CA50 and the actual CA50, the squared error is utilized as the Lyapunov function.

$$V[x(k)] = (y_d - y)^2 \tag{36}$$

With the definition of the Lyapunov function, one can conclude the following:

$$\begin{cases} V[x(k)] = 0, \text{if } y_d - y = 0, \\ V[x(k)] > 0, \forall\ y_d - y \neq 0, \text{and} \\ V[x(k)] \to \infty, \text{if } y_d - y \to \infty. \end{cases} \tag{37}$$

Based on the system equations, the error between the desired output and the actual output can be derived as

$$y_d - y = \alpha(\bar{x}_1(k) - x_1(k)) + \beta(\bar{x}_2(k) - x_2(k)). \tag{38}$$

Substituting Eqn. (38) into Eqn. (36), the Lyapunov function can be captured by

$$V[x(k)] = [\alpha(\bar{x}_1(k) - x_1(k)) + \beta(\bar{x}_2(k) - x_2(k))]^2. \tag{39}$$

Similar to Eqn. (39), Lyapunov function at the $k+1$ cycle can be given as

$$V[x(k+1)] = [\alpha(\bar{x}_1(k+1) - x_1(k+1)) + \beta(\bar{x}_2(k+1) - x_2(k+1))]^2. \tag{40}$$

Once at steady state conditions, the states remain constant or

$$x_1(k+1) = x_1(k), \tag{41}$$

$$x_2(k+1) = x_2(k). \tag{42}$$

Substituting Eqns. (41) and (42) into Eqn. (40), the Lyapunov function at the $k+1$ cycle can be rewritten as

$$V[x(k+1)] = [\alpha(\bar{x}_1(k+1) - x_1(k)) + \beta(\bar{x}_2(k+1) - x_2(k))]^2. \tag{43}$$

Substituting the update equations in Eqns. (32) and (33) into Eqn. (43), the Lyapunov function can be captured by:

$$V[x(k+1)] = \left[ \alpha \left( \bar{x}_1(k) + \frac{0.3\alpha}{\alpha^2 + \beta^2}(y - y_d) - x_1(k) \right) \right. \\ \left. + \beta \left( \bar{x}_2(k) + \frac{0.3\beta}{\alpha^2 + \beta^2}(y - y_d) - x_2(k) \right) \right]^2 \tag{44}$$

Eqn. (44) can be rewritten as

$$V[x(k+1)] = \left[ \alpha(\bar{x}_1(k) - x_1(k)) + \beta(\bar{x}_2(k) - x_2(k)) \right. \\ \left. + \frac{0.3(\alpha^2 + \beta^2)}{\alpha^2 + \beta^2}(y - y_d) \right]^2 \tag{45}$$

and simplified to

$$V[x(k+1)] = [\alpha(\bar{x}_1(k) - x_1(k)) + \beta(\bar{x}_2(k) - x_2(k)) - 0.3(y_d - y)]^2. \tag{46}$$

Substituting Eqn. (38) into Eqn. (46), the Lyapunov function can be expressed as

$$V[x(k+1)] = [y_d - y - 0.3(y_d - y)]^2 \tag{47}$$

and simplified to
$$V[x(k+1)] = 0.49(y_d - y)^2. \tag{48}$$
Therefore, the difference between the Lyapunov function at the $k+1$ cycle and $k$ cycle can be found by subtracting Eqn. (36) from Eqn. (48). This yields
$$V[x(k+1)] - V[x(k)] = -0.51(y_d - y)^2. \tag{49}$$
From Eqn. (49), it can be concluded that:
$$V[x(k+1)] - V[x(k)] < 0, \forall \, y_d - y \neq 0. \tag{50}$$
From Eqns. (37) and Eqn. (50), it can be shown that:
$$\begin{cases} V[x(k)] = 0, \text{if } y_d - y = 0, \\ V[x(k)] > 0, \forall \, y_d - y \neq 0, \\ V[x(k)] \to \infty, \text{if } y_d - y \to \infty \text{ and} \\ V[x(k+1)] - V[x(k)] < 0, \forall \, y_d - y \neq 0. \end{cases} \tag{51}$$
Following the Lyapunov direct method in [39], the output of control system is globally asymptotically stable.